\documentclass[fleqn,10pt]{wlscirep}
\usepackage[utf8]{inputenc}
\usepackage[T1]{fontenc}
\usepackage{rotating}

\usepackage{xcolor}

\RequirePackage[numbers,sort&compress]{natbib}

\graphicspath{ {./graphics/} }

\usepackage{setspace}
\usepackage[flushleft]{threeparttable}

\usepackage{lineno}

\title{The spatial dissemination of COVID-19 and associated socio-economic consequences}

\author[1,2]{Yafei Zhang}
\author[1]{Lin Wang}
\author[2,*]{Jonathan J. H. Zhu}
\author[1,3,*]{Xiaofan Wang}
\affil[1]{Department of Automation, Shanghai Jiao Tong University, and Key Laboratory of System Control and Information Processing, Ministry of Education of China, Shanghai 200240, China}
\affil[2]{Department of Media and Communication, and School of Data Science, City University of Hong Kong, Hong Kong S.A.R., China}
\affil[3]{Department of Automation, Shanghai University, Shanghai 200444, China}

\begin{abstract}
The ongoing coronavirus disease 2019 (COVID-19) pandemic has wreaked havoc worldwide
with millions of lives claimed, human travel restricted, and economic development halted.
Leveraging city-level mobility and case data,
our analysis shows that the spatial dissemination of COVID-19 can be well explained by 
a local diffusion process in the mobility network rather than a global diffusion process,
indicating the effectiveness of the implemented disease prevention and control measures.
Based on the constructed case prediction model,
it's estimated that there could be distinct social consequences if the COVID-19 outbreak occurred in different areas.
During the epidemic control period,
human mobility experienced substantial reductions and the mobility network underwent remarkable local and global structural changes toward containing the spread of COVID-19.
Our work has important implications for the mitigation of disease and the evaluation of the 
socio-economic consequences of COVID-19 on society.

\end{abstract}

\begin{document}

\flushbottom
\maketitle

\thispagestyle{empty}


\section{Introduction}

Severe acute respiratory syndrome coronavirus 2 (SARS-CoV-2), which caused coronavirus disease 2019 (COVID-19), was identified in Wuhan (the provincial capital of Hubei province) in December 2019 and then diffused across mainland China, coinciding with mass human migration during the Spring Festival period \cite{li2020early, chen2020covid}.
Given the migration scale and the position of Wuhan in the national transportation network, combating the dissemination of SARS-CoV-2 became urgent and very challenging.

As the Lunar New Year approaches, a series of disease prevention and control measures were implemented, which effectively contained the evolution of the COVID-19 outbreak in early 2020 \cite{chen2020covid, lai2020effect, pan2020association, kupferschmidt2020can}.
For example, people are encouraged to stay at home in a 14-day nationwide epidemic control period with the coming of the Lunar New Year.
%
After 9 February 2020, the economic reopening was then put in force orderly in most areas due to the notably positive momentum in epidemic control.

Human movements are thought to play a crucial role in shaping the spatio-temporal transmission of infectious diseases \cite{balcan2009multiscale, wesolowski2012quantifying, brockmann2013hidden, jia2020population, kraemer2020effect, mu2021interplay, badr2020association, kissler2020reductions, zhong2020correlation, xu2021multiscale, chinazzi2020effect, li2021understanding, niu2020how, franch2021review}.
To this end, a wealth of studies has been dedicated to investigating the relationship between human mobility and COVID-19 spreading using statistical analysis \cite{jia2020population, kraemer2020effect, tian2020investigation, mu2021interplay, zhong2020correlation, niu2020how, shen2020covid, cheng2020coupled, wei2020examining} and epidemiological modelling \cite{chinazzi2020effect, lai2020effect, chang2021mobility, li2021understanding, huang2021integrated}.
In this paper, we complement these studies by addressing the spatial spread of COVID-19 from the view of network diffusion.
Specifically, using human mobility and case data across more than 360 cities in mainland China, we construct a national human mobility network and assess how the spatial dissemination of COVID-19 is associated with the mobility patterns and what it could be if the COVID-19 outbreak happened in different areas.
Our analysis suggests that the spatial dissemination of COVID-19 in mainland China can be well explained by the human flow from Wuhan and city population
which constitute a local diffusion process in the mobility network,
rather than a global diffusion process
where cities located at central positions are likely to get more infections due to the travel of infected people.
This also indicates the effectiveness of the implemented disease prevention and control measures, where most of the infected people were quarantined or isolated during the epidemic control period, thus largely preventing further transmission to other areas.
Based on the obtained insights, a simple case prediction model is then constructed to estimate potential social consequences if the outbreak occurred in different areas. 
The estimation suggests that where did the COVID-19 occur would play an important role in shaping the spatial prevalence of COVID-19.

"The COVID-19 pandemic is far more than a health crisis: it is affecting societies and economies at their core" \cite{united2020framework}.
The implemented disease prevention and control measures not only significantly changed the course of COVID-19 spreading, but also triggered substantial changes in human mobility and forced reevaluation of social and economic development \cite{united2020framework, imf2020economic, gibbs2020changing, schlosser2020covid, bonaccorsi2020economic}.
Although several valuable attempts have been devoted to this field \cite{bonaccorsi2020economic, nicola2020socio, gibbs2020changing, schlosser2020covid, xiong2020mobile, tong2020short, liu2020evaluating, xu2020analysis, josephson2021socioeconomic, tan2021mobility},
there is still an immense shortage of empirical evaluation of the socio-economic impacts of COVID-19 on society.
Based on the collected human mobility data, this paper further presents an empirical assessment of the social changes in response to COVID-19.
Specifically, we observe a long-lasting reduction of mass migration, where human movements were reduced substantially during the epidemic control period and steadily resumed after the reopening.
The human mobility network experienced striking structural changes as well, with the average path length increasing drastically while the average degree decreasing substantially during the epidemic control period.
As the human mobility network provides the primary pathway along which infectious diseases were transmitted from one city to another, these significant social changes would in turn contribute a lot to combating the spread of COVID-19 \cite{schlosser2020covid, buckee2020aggregated, chang2021mobility}.
Our study helps to understand the spatial dissemination of COVID-19 and could shed light on the modelling of disease spreading and the evaluation of socio-economic consequences in the post-epidemic period.

\begin{figure}[!t]
    \centering
    \includegraphics[width=\hsize]{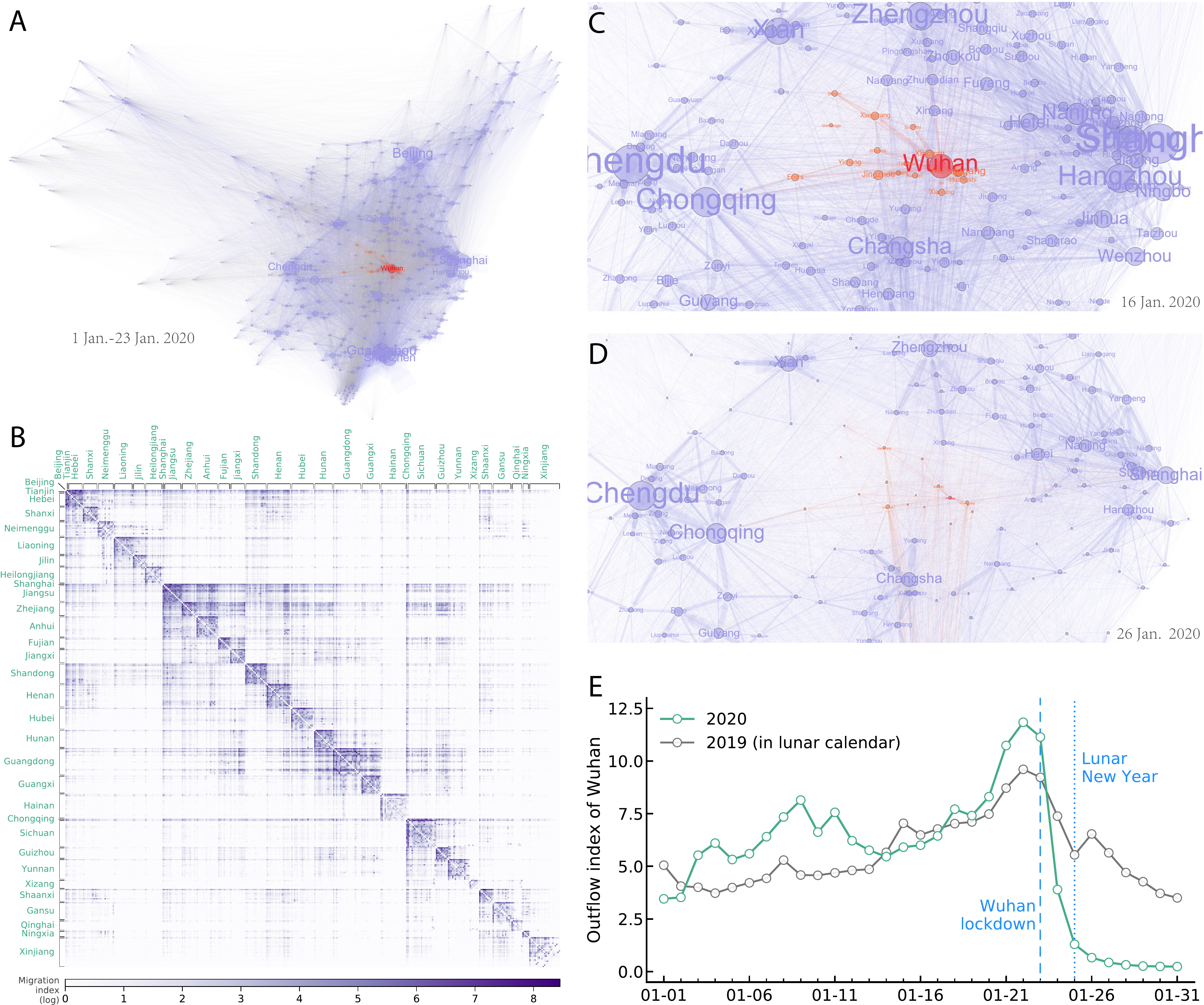}
    \caption{
    \textsf{
    \footnotesize
    \textbf{
    Human mobility network.
    }
   (\textbf{A}) 
   Human mobility network depicted by Baidu Migration Index. 
   Node and label sizes are proportional to the weighted degree of each city in the constructed human mobility network, and edge width is proportional to the volume of human movements.
   (\textbf{B}) 
   Heatmap of the human mobility data corresponding to (\textbf{A}), with cities in the same province (shown in green) placed together.
   For ease of visualization, the raw Baidu Migration Index is multiplied by 100 and then log-transformed by $ln(x+1)$.
   (\textbf{C}) Human mobility network of Hubei (colored in orange with Wuhan highlighted in red) and nearby areas on 16 January 2020.
   (\textbf{D}) Same to (\textbf{C}) but on 26 January 2020.
   (\textbf{E}) Outflow index of Wuhan in January 2020 compared with that in 2019, aligned by the Lunar New Year (which is 25 January in 2020).
    }}
    \label{fig:figure1}
\end{figure}

\section{Results}

\subsection{Human mobility network}

The human mobility data were collected from Baidu Migration platform \cite{baidu2020} which is curated by Chinese search engine Baidu based on its location-based services.
This platform presents relative daily human movements (depicted by the Baidu Migration Index) rather than the exact number of travelers across cities and provinces in mainland China.
We collected the human flow data of 366 cities at the municipal level which cover most of the areas in mainland China.
The national human mobility network is then constructed based on the human movements across cities (see \hyperref[sec:methods]{Methods}
and {Supplementary Material} for details).

Figure \ref{fig:figure1}A illustrates the aggregated human mobility network from 1 January 2020 to 23 January 2020, with nodes representing cities and edges representing the human flow between them.
Cities are placed according to their geographical coordinates, and node and label sizes are proportional to the weighted degree of each city in the constructed mobility network.
Cities in Hubei province and the human migration from them are highlighted in color.
The corresponding human migration data are further presented in Figure \ref{fig:figure1}B, where cities in the same province are placed together and darker colors indicate larger values of human migration.
For ease of visualization, only province names are shown, and provincial capital arrives first in each provincial block.
As shown in the figure, most of the large values are condensed around the diagonal in the migration matrix, which may suggest a clustered structure of human mobility where human movements primarily circulate from one city to another in the same province.

To contain the spread of COVID-19, Wuhan was put on lockdown on 23 January 2020 (two days before the upcoming Lunar New Year).
Shortly, similar epidemic control measures were also implemented in many other cities in Hubei province.
As shown in Figure \ref{fig:figure1}E, the lockdown drastically reduced the population flow from Wuhan to other areas.
For example, compared with last year (2019 in lunar calendar) the human migration from Wuhan dropped about 75\% on the first day (25 January 2020) and 90\% on the third day (27 January 2020) of the Lunar New Year.
Figure \ref{fig:figure1} C-D present two snapshots of the daily human mobility network before (16 January 2020) and after (26 January 2020) the lockdown.
Clearly, the implemented epidemic control measures had effectively cut off the social connections between Hubei and other areas.

\begin{figure}[!tb]
    \centering
    \includegraphics[width=\hsize]{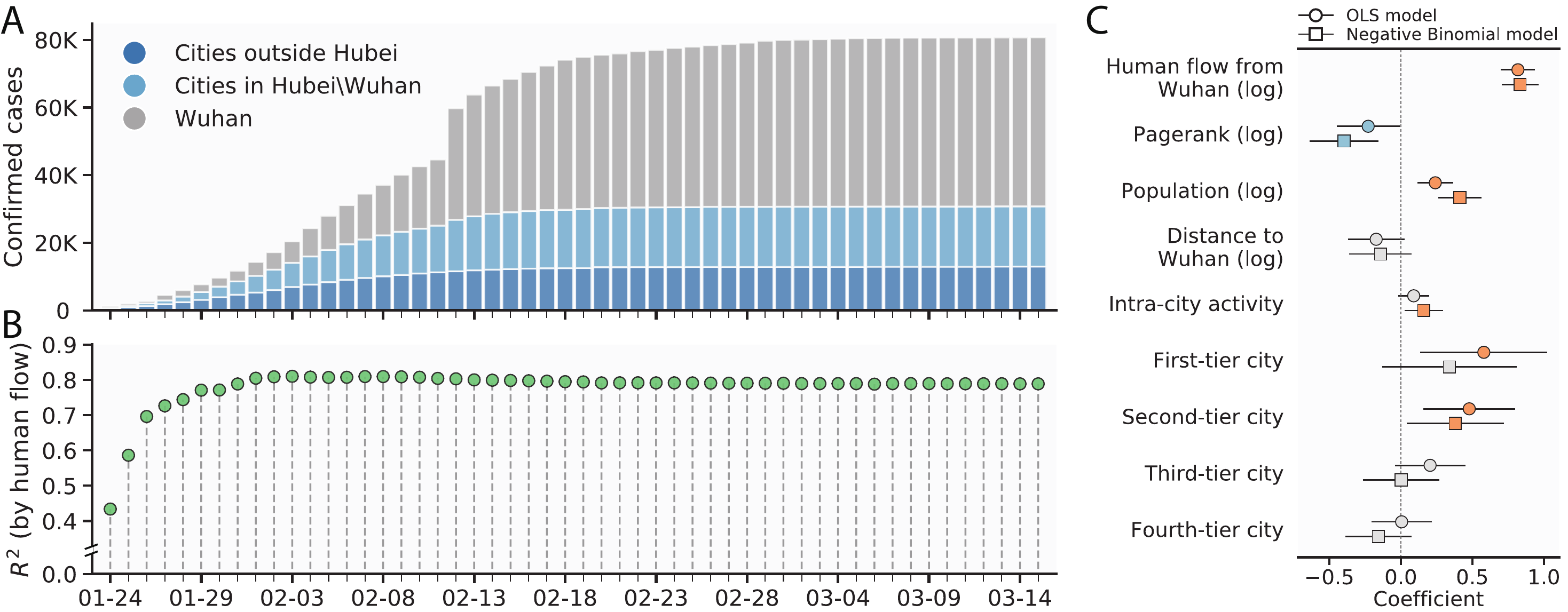}
    \caption{
    \textsf{
    \footnotesize
    \textbf{
    Spatial spread of COVID-19.
    }
    (\textbf{A}) Daily cumulative COVID-19 cases.
    Cities outside of Hubei province are shown in dark blue, cities inside Hubei province (excluding Wuhan) are shown in light blue, and city of Wuhan is shown in grey.
    (\textbf{B}) Human flow from Wuhan explains the spatial distribution of COVID-19 cases.
    The $R^2$ value in each day is obtained by a univariate OLS regression using the number of cumulative cases (log-transformed) of each city on that day as a function of the human flow from Wuhan (log-transformed).
    (\textbf{C}) Estimated coefficients from multiple OLS regression (shown in circles) and Negative Binomial regression (shown in squares) are plotted, with error bars indicating 95\% confidence intervals.
    Estimates whose 95\% confidence intervals do not cross 0 are colored.
    }}
    \label{fig:figure2}
\end{figure}

\subsection{The spatial dissemination of COVID-19}

Catalyzed by the annual Spring Festival Travel Rush (involves as many as three billion trips in a 40-day period in 2019) and the improved clinical testing capacity, the number of confirmed COVID-19 cases was escalating with the arrival of the Lunar New Year (Figure \ref{fig:figure2}A).
Consistent with previous studies \cite{jia2020population, kraemer2020effect, mu2021interplay, niu2020how, wei2020examining},
we find that the spatial prevalence of COVID-19 in mainland China can be well explained (measured by $R^2$) by the human flow from Wuhan (1-23 January 2020) (Figure \ref{fig:figure2}B).
For a given date, the $R^2$ value is obtained by a univariate Ordinary Least Squares (OLS) regression using the number of cumulative cases (log-transformed) on that day as a function of human flow from Wuhan (log-transformed)
(see {Supplementary Note 2} for further details).
Specially, we achieve a $R^2$ value of approximately 0.8 since 31 January 2020.




We further adopt multivariate regression model and incorporate more city-specific factors in the analysis,
including the global centrality of a city in the mobility network (measured by Pagerank \cite{page1999pagerank, langville2005survey}),
city population,
the spatial distance to Wuhan,
intra-city activity intensity (provided by Baidu), 
and city tier.
From a network perspective, 
the human flow from Wuhan captures a local diffusion process of COVID-19--from Wuhan to neighboring areas--in the mobility network, 
while Pagerank would denote a global network diffusion process that involves multi-step transmissions across the network (
Figure S3,
{Supplementary Note 2}).
Therefore, the direct comparison between the human flow from Wuhan and the global centrality of a city (Pagerank) would be able to answer the following question:
which diffusion process dominates the spatial dissemination of COVID-19, local or global network diffusion?
Using the variables described above, both OLS and Negative Binomial regression models are adopted in the analysis
(see {Supplementary Material} for further details).

Figure \ref{fig:figure2}C illustrates the estimated coefficients for each variable in predicting the spatial distribution of accumulative COVID-19 cases on 9 February 2020.
Specifically, we find consistent evidence that both human flow from Wuhan and city population act as significant and positive predictors ($p<0.001$) in the case prediction
(Figure \ref{fig:figure2}C).
In other words, cities with larger volumes of human migration from Wuhan and more population are likely to get more infections.
More importantly, classic complex network spreading theory would hypothesize that cities located at central positions in the mobility network are generally vulnerable to infectious diseases.
However, our study reveals that although the global network centrality of a city (measured by Pagerank) is positively correlated with the number of confirmed cases
(Spearman's $r_s=0.6698$, $p<0.001$),
once the human flow from Wuhan and city population are controlled in the regression, the positive role of the global network centrality in the prediction of cumulative COVID-19 cases
disappears
(Figure \ref{fig:figure2}C and {Supplementary Note 2}).
The finding suggests that the spatial dissemination of COVID-19 in mainland China can be well explained by a local network diffusion process which goes only one step further from the outbreak area in the mobility network, rather than a global network diffusion process.
It also implies the effectiveness of the implemented control measures where most of the infected people were quarantined and isolated during the epidemic control period, thereby largely preventing further transmission to other areas.
In other words, without effective control measures,
a global network diffusion process of COVID-19 may be discovered, and cities located at central positions may have much more people infected due to the migration of the infected across areas.

\begin{figure}[!tb]
    \centering
    \includegraphics[width=16cm]{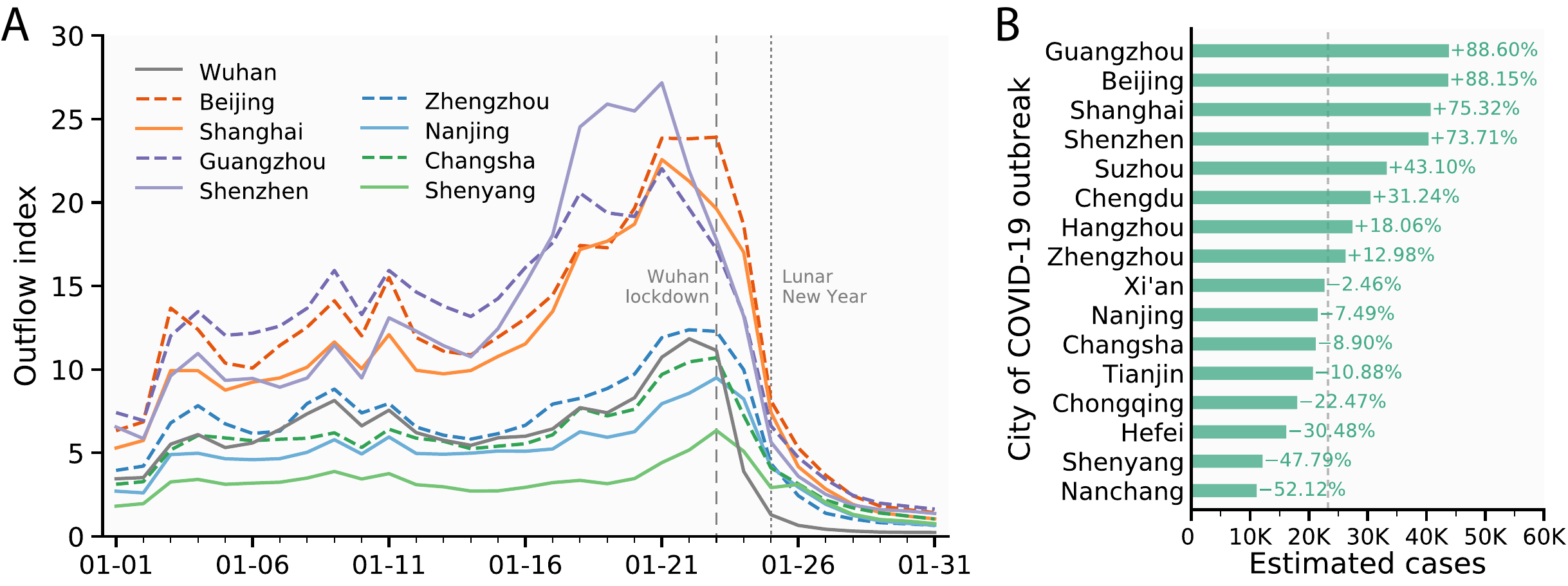}
    \caption{
    \textsf{
    \footnotesize
    \textbf{
    COVID-19 outbreak in different areas.
    }
    (\textbf{A})
    Outflow index of nine example cities.
    (\textbf{B}) Estimated accumulative COVID-19 cases if the outbreak happened in different areas.
    The vertical dashed line indicates the actual number of confirmed cases (excluding Wuhan) on 9 February 2020 (serves as the baseline), and the number after each bar indicates the relative change of confirmed cases compared with the baseline.
    }}
    \label{fig:figure3}
\end{figure}

\subsection{COVID-19 outbreak in different areas}

Based on the obtained insights above, we estimate what it could be had the COVID-19 outbreak occurred in different areas.
We focus on several key factors that help to predict the prevalence of COVID-19, including
the human flow from and the distance to the outbreak city, city population, and intra-city activity intensity.
Figure \ref{fig:figure3}A presents the outflow index of nine example cities in January 2020, where some cities (e.g., Beijing, Shanghai, and Guangzhou) underwent much more population outflow than Wuhan while some others (e.g., Changsha and Shenyang) underwent relatively less population outflow.
For the outbreak in Wuhan, we construct a Negative Binomial regression model with the cumulative number of cases on 9 February 2020 set as the dependent variable and the above key factors set as the independent variables.
After that, we obtain the spatial spread pattern of COVID-19 depicted by these factors.
Suppose that the control measures and the spatial spread pattern remain the same.
Based on the constructed model, the spatial prevalence of COVID-19 can be roughly estimated when the outbreak area changes
(see {Supplementary Note 3} for further details).

Figure \ref{fig:figure3}B illustrates the estimated cumulative cases (excluding the outbreak area) as of 9 February 2020, varying with the outbreak area.
The vertical dashed line indicates the actual cumulative number of confirmed cases in cities other than Wuhan on 9 February 2020 (which is 23,236) and serves as the baseline.
Compared with the baseline, the relative change of cumulative cases for each outbreak area is shown in percentage.
As shown in the figure, if the COVID-19 outbreak happened in cities like Beijing and Guangzhou, the number of confirmed cases could be nearly doubled,
but if the outbreak occurred in cities like Shenyang and Nanchang, the number of confirmed cases could be reduced by nearly half.
This also suggests that where did the outbreak occur could play an important role in the spatial dissemination of COVID-19, which may have meaningful implications for the prevention of infectious diseases in the future.

\subsection{Social changes}

\begin{figure}[!tb]
    \centering
    \includegraphics[width=\hsize]{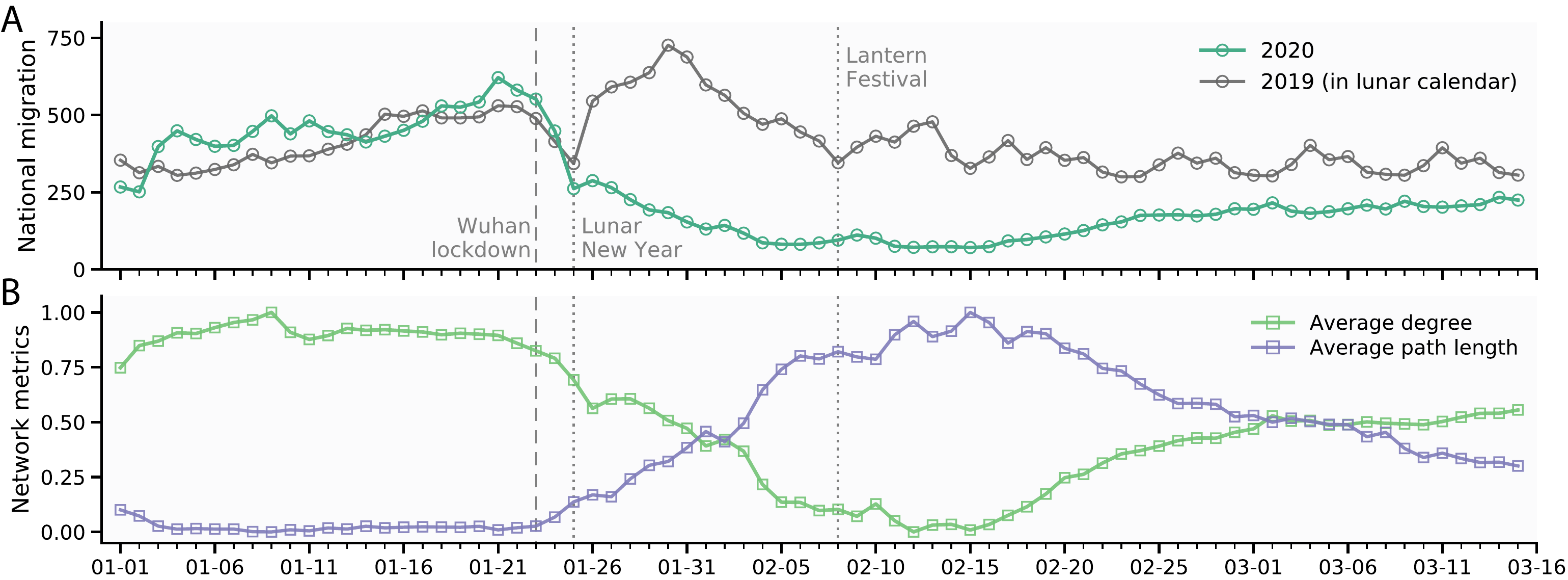}
    \caption{
    \textsf{
    \footnotesize
    \textbf{
    Human mobility during the COVID-19.
    }
    (\textbf{A}) Daily national migration.
    (\textbf{B}) Mobility network changes in terms of average degree and average path length (normalized to the range [0, 1]).
    }}
    \label{fig:figure4}
\end{figure}

After the implementation of a series of epidemic control measures, human mobility 
underwent
striking changes.
Usually, we would expect a recovery of human movements since the second day of the Lunar New Year.
However, due to the outbreak of COVID-19, the national migration witnessed drastic and long-lasting shrinkage since the Lunar New Year
(Figure \ref{fig:figure4}A).
For example, on the sixth day of the Lunar New Year (30 January 2020), the national migration scale dropped by nearly three quarters compared with last year.
Specifically, instead of a travel surge immediately after the Lunar New Year, we observe that
the national migration scale gradually decreased until the coming of the Lantern Festival (close to the reopening).
After the economic reopening was put in force orderly, the national migration steadily resumed afterwards.

We also observe remarkable local and global structural changes of the mobility network
(Figure \ref{fig:figure4}B).
First, after the implementation of a series of epidemic control measures, the average degree of the mobility network endured notable reductions before the economic reopening.
This local structural change would reduce the connectivity of the mobility network and was able to prevent the spread of virus across areas.
Moreover, the average path length of the mobility network experienced substantial increases as well during the epidemic control period.
This global structural change would largely reduce the reachability of each area in the mobility network and was able to delay the spread of virus from one place to another.
Taken together, these mobility changes during the control period would, in turn, contribute to the mitigation of infectious diseases \cite{pastor2015epidemic, schlosser2020covid}.
After the reopening, especially after 15 February 2020, we observe a steady recovery of the network connectivity and reachability, which also indicates the lifting of travel restrictions across the country.

\section{Discussion}

The COVID-19 pandemic is a serious crisis and a daunting challenge for the entire world.
In this paper, our analysis shows that the spatial dissemination of COVID-19 in mainland China can be well explained by a local network diffusion process rather than a global network diffusion process,
which implies the effectiveness of the implemented epidemic control measures.
It's also estimated that there could be very different social consequences if the COVID-19 outbreak area changes, which may have meaningful implications for future epidemic prevention and control.
We also note a remarkable reduction of human movements during the epidemic control period,
with significant structural changes of the human mobility network toward containing the spread of COVID-19.
In summary, our work contributes to a further understanding of how human mobility data and network analysis can be utilized to address the spread of infectious diseases and paves a way for the application of data analytics in preventing and containing an epidemic.

Our work has several limitations as well.
First, we emphasize that most of our conclusions are drawn upon correlation studies based on observational data, thereby not reflecting causality sufficiently.
Second, the mobility data we adopted here are obtained from Baidu based on its location-based services, but for those who don't adopt such services, we are not able to incorporate their movements in the current study.
Other sources of mobility data are thus needed to enhance the analysis.
Third, due to the lack of accurate timestamps of human movements,
we don't exactly know the departure and arrival time of each travel trajectory.
Therefore, there may exist travel delay issues in the human mobility data.
For example, some people may depart from a city in one day but arrive at the destination in the following days.
In addition, this paper mainly investigates the spatial dissemination of COVID-19 in mainland China, but whether the proposed approach applies in other areas or other kinds of infectious diseases still needs further exploration in the future.

\section{Methods}

\label{sec:methods}

\subsection*{Data}

The human mobility data were sourced from the Baidu Migration platform \cite{baidu2020} based on Baidu's location-based services.
As the dominant search engine in China, Baidu has nearly 189 million daily active users
and responses to more than 120 billion daily location service requests.
Similar to previous studies \cite{lai2020effect, kraemer2020effect}, the mobility data
don't indicate the absolute number of recorded trips but reflect the relative movements of people using Baidu's location-based services.
We collected daily inter- and intra-city mobility data across 366 cities from 1 January to 15 March in 2020 and the corresponding period in 2019 (aligned by the Lunar New Year).
For inter-city activity in 2019, only aggregated inflow and outflow data were provided for each city.
The COVID-19 data were obtained from the daily case report released by the Health Commission of each province and NetEase News \cite{netease2020}, a professional media platform that provides timely updates and serves as a supplementary source in our study.
The population of each city
was collected from 
the National Economic and Social Development Statistical Bulletin 2019.
The spatial distance between two cities was obtained by their geodesic distance based on their latitude and longitude geographical coordinates.

\subsection*{Network analysis}

We adopt Pagerank \cite{page1999pagerank, langville2005survey}, a classic global network centrality measure, to quantify how important a city is located in the mobility network.
In practice, the human flow volume between two cities is used as the weight in the calculation of Pagerank.
The average degree measures the local connectivity of the mobility network and can be simply calculated using the average number of incoming and outgoing links of each node in the mobility network.
In the context of human migration, two cities are said to be close to each other if they share a large volume of human flow \cite{schlosser2020covid}.
As such, we use the inverse of the human flow volume to denote the ``network distance'' of 
two cities along each edge, based on which the shortest path length from one city to another is calculated.
The average path length of the mobility network is thus obtained by averaging the shortest path length of all pairs of nodes.
In practice, these network metrics were obtained using Python package \texttt{networkx}.
Given a vector $V$ comprising a list of quantities, the element of $V$ is normalized as
$V_{i}=\left(V_{i}-V_{\min }\right) /\left(V_{\max }-V_{\min }\right)$,
where $V_{\max }$ and $V_{\min }$ are the maximum and minimum values of $V$,
respectively.

\subsection*{Statistical analysis}
Most of the data processing was done by Python package \texttt{pandas} and R package \texttt{dplyr}.
Spearman rank correlation was performed by Python package \texttt{scipy};
OLS regression analysis was performed by Python package \texttt{statsmodel}
and R function \texttt{lm};
Negative Binomial regression was performed by 
R package \texttt{MASS}
and 
Python package \texttt{statsmodel}.
Further details on statistical analysis can be found in {Supplementary Material}.

\section*{Supplementary Material}

The Supplementary Material is available at:
\noindent
\url{https://doi.org/10.6084/m9.figshare.19145174.v1}.

\bibliography{sample}

\begin{thebibliography}{10}
\expandafter\ifx\csname url\endcsname\relax
  \def\url#1{\texttt{#1}}\fi
\expandafter\ifx\csname urlprefix\endcsname\relax\def\urlprefix{URL }\fi
\providecommand{\bibinfo}[2]{#2}
\providecommand{\eprint}[2][]{\url{#2}}

\bibitem{li2020early}
\bibinfo{author}{Li, Q.} \emph{et~al.}
\newblock \bibinfo{title}{Early transmission dynamics in {Wuhan}, {China}, of
  novel coronavirus-infected pneumonia}.
\newblock \emph{\bibinfo{journal}{New England Journal of Medicine}}
  \textbf{\bibinfo{volume}{382}}, \bibinfo{pages}{1199--1207}
  (\bibinfo{year}{2020}).

\bibitem{chen2020covid}
\bibinfo{author}{Chen, S.}, \bibinfo{author}{Yang, J.}, \bibinfo{author}{Yang,
  W.}, \bibinfo{author}{Wang, C.} \& \bibinfo{author}{B{\"a}rnighausen, T.}
\newblock \bibinfo{title}{{COVID-19} control in {China} during mass population
  movements at new year}.
\newblock \emph{\bibinfo{journal}{The Lancet}} \textbf{\bibinfo{volume}{395}},
  \bibinfo{pages}{764--766} (\bibinfo{year}{2020}).

\bibitem{lai2020effect}
\bibinfo{author}{Lai, S.} \emph{et~al.}
\newblock \bibinfo{title}{Effect of non-pharmaceutical interventions to contain
  {COVID-19} in {China}}.
\newblock \emph{\bibinfo{journal}{Nature}} \textbf{\bibinfo{volume}{585}},
  \bibinfo{pages}{410--413} (\bibinfo{year}{2020}).

\bibitem{pan2020association}
\bibinfo{author}{Pan, A.} \emph{et~al.}
\newblock \bibinfo{title}{Association of public health interventions with the
  epidemiology of the {COVID-19} outbreak in {Wuhan}, {China}}.
\newblock \emph{\bibinfo{journal}{JAMA}} \textbf{\bibinfo{volume}{323}},
  \bibinfo{pages}{1915--1923} (\bibinfo{year}{2020}).

\bibitem{kupferschmidt2020can}
\bibinfo{author}{Kupferschmidt, K.} \& \bibinfo{author}{Cohen, J.}
\newblock \bibinfo{title}{Can {China}{\textquoteright}s {COVID-19} strategy
  work elsewhere?}
\newblock \emph{\bibinfo{journal}{Science}} \textbf{\bibinfo{volume}{367}},
  \bibinfo{pages}{1061--1062} (\bibinfo{year}{2020}).

\bibitem{balcan2009multiscale}
\bibinfo{author}{Balcan, D.} \emph{et~al.}
\newblock \bibinfo{title}{Multiscale mobility networks and the spatial
  spreading of infectious diseases}.
\newblock \emph{\bibinfo{journal}{Proceedings of the National Academy of
  Sciences}} \textbf{\bibinfo{volume}{106}}, \bibinfo{pages}{21484--21489}
  (\bibinfo{year}{2009}).

\bibitem{wesolowski2012quantifying}
\bibinfo{author}{Wesolowski, A.} \emph{et~al.}
\newblock \bibinfo{title}{Quantifying the impact of human mobility on malaria}.
\newblock \emph{\bibinfo{journal}{Science}} \textbf{\bibinfo{volume}{338}},
  \bibinfo{pages}{267--270} (\bibinfo{year}{2012}).

\bibitem{brockmann2013hidden}
\bibinfo{author}{Brockmann, D.} \& \bibinfo{author}{Helbing, D.}
\newblock \bibinfo{title}{The hidden geometry of complex, network-driven
  contagion phenomena}.
\newblock \emph{\bibinfo{journal}{Science}} \textbf{\bibinfo{volume}{342}},
  \bibinfo{pages}{1337--1342} (\bibinfo{year}{2013}).

\bibitem{jia2020population}
\bibinfo{author}{Jia, J.~S.} \emph{et~al.}
\newblock \bibinfo{title}{Population flow drives spatio-temporal distribution
  of {COVID-19} in {China}}.
\newblock \emph{\bibinfo{journal}{Nature}} \textbf{\bibinfo{volume}{582}},
  \bibinfo{pages}{389--395} (\bibinfo{year}{2020}).

\bibitem{kraemer2020effect}
\bibinfo{author}{Kraemer, M.~U.} \emph{et~al.}
\newblock \bibinfo{title}{The effect of human mobility and control measures on
  the {COVID-19} epidemic in {China}}.
\newblock \emph{\bibinfo{journal}{Science}} \textbf{\bibinfo{volume}{368}},
  \bibinfo{pages}{493--497} (\bibinfo{year}{2020}).

\bibitem{mu2021interplay}
\bibinfo{author}{Mu, X.}, \bibinfo{author}{Yeh, A. G.-O.} \&
  \bibinfo{author}{Zhang, X.}
\newblock \bibinfo{title}{The interplay of spatial spread of {COVID-19} and
  human mobility in the urban system of {China} during the {Chinese New Year}}.
\newblock \emph{\bibinfo{journal}{Environment and Planning B: Urban Analytics
  and City Science}} \textbf{\bibinfo{volume}{48}}, \bibinfo{pages}{1955--1971}
  (\bibinfo{year}{2021}).

\bibitem{badr2020association}
\bibinfo{author}{Badr, H.~S.} \emph{et~al.}
\newblock \bibinfo{title}{Association between mobility patterns and {COVID-19}
  transmission in the {USA}: a mathematical modelling study}.
\newblock \emph{\bibinfo{journal}{The Lancet Infectious Diseases}}
  \textbf{\bibinfo{volume}{20}}, \bibinfo{pages}{1247--1254}
  (\bibinfo{year}{2020}).

\bibitem{kissler2020reductions}
\bibinfo{author}{Kissler, S.~M.} \emph{et~al.}
\newblock \bibinfo{title}{Reductions in commuting mobility correlate with
  geographic differences in {SARS-CoV-2} prevalence in {New York City}}.
\newblock \emph{\bibinfo{journal}{Nature Communications}}
  \textbf{\bibinfo{volume}{11}}, \bibinfo{pages}{4674} (\bibinfo{year}{2020}).

\bibitem{zhong2020correlation}
\bibinfo{author}{Zhong, P.}, \bibinfo{author}{Guo, S.} \&
  \bibinfo{author}{Chen, T.}
\newblock \bibinfo{title}{{Correlation between travellers departing from Wuhan
  before the Spring Festival and subsequent spread of COVID-19 to all provinces
  in China}}.
\newblock \emph{\bibinfo{journal}{Journal of Travel Medicine}}
  \textbf{\bibinfo{volume}{27}} (\bibinfo{year}{2020}).
\newblock
  \eprint{https://academic.oup.com/jtm/article-pdf/27/3/taaa036/33226371/taaa036.pdf}.

\bibitem{xu2021multiscale}
\bibinfo{author}{Xu, X.-K.}, \bibinfo{author}{Wang, L.} \&
  \bibinfo{author}{Pei, S.}
\newblock \bibinfo{title}{{Multiscale mobility explains differential
  associations between the gross domestic product and COVID-19 transmission in
  Chinese cities}}.
\newblock \emph{\bibinfo{journal}{Journal of Travel Medicine}}
  \textbf{\bibinfo{volume}{28}} (\bibinfo{year}{2021}).
\newblock
  \eprint{https://academic.oup.com/jtm/article-pdf/28/2/taaa236/36338275/taaa236.pdf}.

\bibitem{chinazzi2020effect}
\bibinfo{author}{Chinazzi, M.} \emph{et~al.}
\newblock \bibinfo{title}{The effect of travel restrictions on the spread of
  the 2019 novel coronavirus ({COVID-19}) outbreak}.
\newblock \emph{\bibinfo{journal}{Science}} \textbf{\bibinfo{volume}{368}},
  \bibinfo{pages}{395--400} (\bibinfo{year}{2020}).

\bibitem{li2021understanding}
\bibinfo{author}{Li, T.}, \bibinfo{author}{Luo, J.} \& \bibinfo{author}{Huang,
  C.}
\newblock \bibinfo{title}{Understanding small {Chinese} cities as {COVID-19}
  hotspots with an urban epidemic hazard index}.
\newblock \emph{\bibinfo{journal}{Scientific Reports}}
  \textbf{\bibinfo{volume}{11}}, \bibinfo{pages}{14663} (\bibinfo{year}{2021}).

\bibitem{niu2020how}
\bibinfo{author}{Niu, X.}, \bibinfo{author}{Yue, Y.}, \bibinfo{author}{Zhou,
  X.} \& \bibinfo{author}{Zhang, X.}
\newblock \bibinfo{title}{How urban factors affect the spatiotemporal
  distribution of infectious diseases in addition to intercity population
  movement in {China}}.
\newblock \emph{\bibinfo{journal}{ISPRS International Journal of
  Geo-Information}} \textbf{\bibinfo{volume}{9}}, \bibinfo{pages}{615}
  (\bibinfo{year}{2020}).

\bibitem{franch2021review}
\bibinfo{author}{Franch-Pardo, I.}, \bibinfo{author}{Desjardins, M.~R.},
  \bibinfo{author}{Barea-Navarro, I.} \& \bibinfo{author}{Cerdà, A.}
\newblock \bibinfo{title}{A review of gis methodologies to analyze the dynamics
  of covid-19 in the second half of 2020}.
\newblock \emph{\bibinfo{journal}{Transactions in GIS}}
  \textbf{\bibinfo{volume}{25}}, \bibinfo{pages}{2191--2239}
  (\bibinfo{year}{2021}).

\bibitem{tian2020investigation}
\bibinfo{author}{Tian, H.} \emph{et~al.}
\newblock \bibinfo{title}{An investigation of transmission control measures
  during the first 50 days of the {COVID-19} epidemic in {China}}.
\newblock \emph{\bibinfo{journal}{Science}} \textbf{\bibinfo{volume}{368}},
  \bibinfo{pages}{638--642} (\bibinfo{year}{2020}).

\bibitem{shen2020covid}
\bibinfo{author}{Shen, J.}
\newblock \bibinfo{title}{{COVID-19} and inter-provincial migration in
  {China}}.
\newblock \emph{\bibinfo{journal}{Eurasian Geography and Economics}}
  \textbf{\bibinfo{volume}{61}}, \bibinfo{pages}{620--626}
  (\bibinfo{year}{2020}).

\bibitem{cheng2020coupled}
\bibinfo{author}{Cheng, C.} \emph{et~al.}
\newblock \bibinfo{title}{The coupled impact of emergency responses and
  population flows on the {COVID-19} pandemic in {China}}.
\newblock \emph{\bibinfo{journal}{GeoHealth}} \textbf{\bibinfo{volume}{4}},
  \bibinfo{pages}{e2020GH000332} (\bibinfo{year}{2020}).

\bibitem{wei2020examining}
\bibinfo{author}{Wei, S.} \& \bibinfo{author}{Wang, L.}
\newblock \bibinfo{title}{Examining the population flow network in {China} and
  its implications for epidemic control based on {Baidu} migration data}.
\newblock \emph{\bibinfo{journal}{Humanities and Social Sciences
  Communications}} \textbf{\bibinfo{volume}{7}}, \bibinfo{pages}{145}
  (\bibinfo{year}{2020}).

\bibitem{chang2021mobility}
\bibinfo{author}{Chang, S.} \emph{et~al.}
\newblock \bibinfo{title}{Mobility network models of {COVID-19} explain
  inequities and inform reopening}.
\newblock \emph{\bibinfo{journal}{Nature}} \textbf{\bibinfo{volume}{589}},
  \bibinfo{pages}{82--87} (\bibinfo{year}{2021}).

\bibitem{huang2021integrated}
\bibinfo{author}{Huang, B.} \emph{et~al.}
\newblock \bibinfo{title}{Integrated vaccination and physical distancing
  interventions to prevent future {COVID-19} waves in {Chinese} cities}.
\newblock \emph{\bibinfo{journal}{Nature Human Behaviour}}
  \textbf{\bibinfo{volume}{5}}, \bibinfo{pages}{695--705}
  (\bibinfo{year}{2021}).

\bibitem{united2020framework}
\bibinfo{author}{{United Nations}}.
\newblock \bibinfo{title}{A {UN} framework for the immediate socio-economic
  response to {COVID-19}} (\bibinfo{year}{April 2020}).
\newblock
  \bibinfo{note}{\url{https://unsdg.un.org/resources/un-framework-immediate-socio-economic-response-covid-19}}.

\bibitem{imf2020economic}
\bibinfo{author}{{International Monetary Fund}}.
\newblock \bibinfo{title}{World economic outlook} (\bibinfo{year}{April 2020}).
\newblock
  \bibinfo{note}{\url{https://www.imf.org/en/Publications/WEO/Issues/2020/04/14/weo-april-2020}}.

\bibitem{gibbs2020changing}
\bibinfo{author}{Gibbs, H.} \emph{et~al.}
\newblock \bibinfo{title}{Changing travel patterns in {China} during the early
  stages of the {COVID-19} pandemic}.
\newblock \emph{\bibinfo{journal}{Nature Communications}}
  \textbf{\bibinfo{volume}{11}}, \bibinfo{pages}{5012} (\bibinfo{year}{2020}).

\bibitem{schlosser2020covid}
\bibinfo{author}{Schlosser, F.} \emph{et~al.}
\newblock \bibinfo{title}{{COVID-19} lockdown induces disease-mitigating
  structural changes in mobility networks}.
\newblock \emph{\bibinfo{journal}{Proceedings of the National Academy of
  Sciences}} \textbf{\bibinfo{volume}{117}}, \bibinfo{pages}{32883--32890}
  (\bibinfo{year}{2020}).

\bibitem{bonaccorsi2020economic}
\bibinfo{author}{Bonaccorsi, G.} \emph{et~al.}
\newblock \bibinfo{title}{Economic and social consequences of human mobility
  restrictions under {COVID-19}}.
\newblock \emph{\bibinfo{journal}{Proceedings of the National Academy of
  Sciences}} \textbf{\bibinfo{volume}{117}}, \bibinfo{pages}{15530--15535}
  (\bibinfo{year}{2020}).

\bibitem{nicola2020socio}
\bibinfo{author}{Nicola, M.} \emph{et~al.}
\newblock \bibinfo{title}{The socio-economic implications of the coronavirus
  pandemic ({COVID-19}): A review}.
\newblock \emph{\bibinfo{journal}{International Journal of Surgery}}
  \textbf{\bibinfo{volume}{78}}, \bibinfo{pages}{185--193}
  (\bibinfo{year}{2020}).

\bibitem{xiong2020mobile}
\bibinfo{author}{Xiong, C.} \emph{et~al.}
\newblock \bibinfo{title}{Mobile device location data reveal human mobility
  response to state-level stay-at-home orders during the {COVID-19} pandemic in
  the {USA}}.
\newblock \emph{\bibinfo{journal}{Journal of the Royal Society Interface}}
  \textbf{\bibinfo{volume}{17}}, \bibinfo{pages}{20200344}
  (\bibinfo{year}{2020}).

\bibitem{tong2020short}
\bibinfo{author}{Tong, Y.}, \bibinfo{author}{Ma, Y.} \& \bibinfo{author}{Liu,
  H.}
\newblock \bibinfo{title}{The short-term impact of {COVID-19} epidemic on the
  migration of {Chinese} urban population and the evaluation of {Chinese} urban
  resilience}.
\newblock \emph{\bibinfo{journal}{Acta Geographica Sinica}}
  \textbf{\bibinfo{volume}{75}}, \bibinfo{pages}{2505--2520}
  (\bibinfo{year}{2020}).

\bibitem{liu2020evaluating}
\bibinfo{author}{Liu, H.}, \bibinfo{author}{Fang, C.} \& \bibinfo{author}{Gao,
  Q.}
\newblock \bibinfo{title}{Evaluating the real-time impact of {COVID-19} on
  cities: {China} as a case study}.
\newblock \emph{\bibinfo{journal}{Complexity}} \textbf{\bibinfo{volume}{2020}},
  \bibinfo{pages}{8855521} (\bibinfo{year}{2020}).

\bibitem{xu2020analysis}
\bibinfo{author}{Xu, X.}, \bibinfo{author}{Wang, S.}, \bibinfo{author}{Dong,
  J.}, \bibinfo{author}{Shen, Z.} \& \bibinfo{author}{Xu, S.}
\newblock \bibinfo{title}{An analysis of the domestic resumption of social
  production and life under the {COVID-19} epidemic}.
\newblock \emph{\bibinfo{journal}{PloS one}} \textbf{\bibinfo{volume}{15}},
  \bibinfo{pages}{e0236387} (\bibinfo{year}{2020}).

\bibitem{josephson2021socioeconomic}
\bibinfo{author}{Josephson, A.}, \bibinfo{author}{Kilic, T.} \&
  \bibinfo{author}{Michler, J.~D.}
\newblock \bibinfo{title}{Socioeconomic impacts of {COVID-19} in low-income
  countries}.
\newblock \emph{\bibinfo{journal}{Nature Human Behaviour}}
  \textbf{\bibinfo{volume}{5}}, \bibinfo{pages}{557--565}
  (\bibinfo{year}{2021}).

\bibitem{tan2021mobility}
\bibinfo{author}{Tan, S.} \emph{et~al.}
\newblock \bibinfo{title}{{Mobility in {China}, 2020: {A} tale of four
  phases}}.
\newblock \emph{\bibinfo{journal}{National Science Review}}
  (\bibinfo{year}{2021}).
\newblock \urlprefix\url{https://doi.org/10.1093/nsr/nwab148}.

\bibitem{buckee2020aggregated}
\bibinfo{author}{Buckee, C.~O.} \emph{et~al.}
\newblock \bibinfo{title}{Aggregated mobility data could help fight
  {COVID-19}.}
\newblock \emph{\bibinfo{journal}{Science}} \textbf{\bibinfo{volume}{368}},
  \bibinfo{pages}{145--146} (\bibinfo{year}{2020}).

\bibitem{baidu2020}
\bibinfo{title}{Baidu {Migration} [in {Chinese}]}.
\newblock \bibinfo{howpublished}{\url{https://qianxi.baidu.com/2020/}}.
\newblock \bibinfo{note}{Accessed: June 2020}.

\bibitem{page1999pagerank}
\bibinfo{author}{Page, L.}, \bibinfo{author}{Brin, S.},
  \bibinfo{author}{Motwani, R.} \& \bibinfo{author}{Winograd, T.}
\newblock \bibinfo{title}{The {PageRank} citation ranking: Bringing order to
  the web.}
\newblock \bibinfo{type}{Tech. Rep.}, \bibinfo{institution}{Stanford InfoLab}
  (\bibinfo{year}{1999}).

\bibitem{langville2005survey}
\bibinfo{author}{Langville, A.~N.} \& \bibinfo{author}{Meyer, C.~D.}
\newblock \bibinfo{title}{A survey of eigenvector methods for web information
  retrieval}.
\newblock \emph{\bibinfo{journal}{SIAM Review}} \textbf{\bibinfo{volume}{47}},
  \bibinfo{pages}{135--161} (\bibinfo{year}{2005}).

\bibitem{pastor2015epidemic}
\bibinfo{author}{Pastor-Satorras, R.}, \bibinfo{author}{Castellano, C.},
  \bibinfo{author}{Van~Mieghem, P.} \& \bibinfo{author}{Vespignani, A.}
\newblock \bibinfo{title}{Epidemic processes in complex networks}.
\newblock \emph{\bibinfo{journal}{Reviews of Modern Physics}}
  \textbf{\bibinfo{volume}{87}}, \bibinfo{pages}{925--979}
  (\bibinfo{year}{2015}).

\bibitem{netease2020}
\bibinfo{title}{{NetEase} {News} [in {Chinese}]}.
\newblock
  \bibinfo{howpublished}{\url{https://wp.m.163.com/163/page/news/virus_report/index.html}}.
\newblock \bibinfo{note}{Accessed: August 2020}.

\end{thebibliography}


\end{document}